\documentstyle[titlepage,12pt]{article}

\def\be{\begin{equation}}
\def\bea{\begin{eqnarray}}

\def\S{{\bf S}}
\def\H{{\bf H}}
\def\C{{\bf C}}
\def\Sbar{{\bf \bar{S}}}
\def\Hbar{{\bf \bar{H}}}
\def\Cbar{{\bf \bar{C}}}
\def\ee{\end{equation}}
\def\eea{\end{eqnarray}}

\newcount\sectionnumber
\def\sect
{\global\equationnumber=0
\global\advance\sectionnumber by 1
\the\sectionnumber . }
\newcount\equationnumber
\def   \num
{\eqno{\global\advance\equationnumber by 1
\left(\the\sectionnumber .\the\equationnumber \right)}}

\begin{document}
\title{New Types of Thermodynamics from\\
$(1+1)$-Dimensional Black Holes}
\author{P.T. Landsberg\\
Dept. of Mathematical Sciences\\
University of Southampton\\
Southampton
United Kingdom SO9 5NH \\
and \\
R.B. Mann \\
Department of Physics \\
University of Waterloo \\
Waterloo, Ontario \\
N2L 3G1}

\date{July 6, 1993\\
WATPHYS TH-93/05}
\maketitle

\begin{abstract}
For normal thermodynamic systems superadditivity $\S$, homogeneity $\H$ and
concavity $\C$ of the entropy hold, whereas for $(3+1)$-dimensional black
holes the latter two properties are violated. We show that
$(1+1)$-dimensional black holes exhibit  qualitatively new types of
thermodynamic behaviour, discussed here for the first time,  in which $\C$
always holds, $\H$ is always violated and $\S$ may or may not be violated,
depending of the magnitude of the black hole mass.  Hence it is now seen
that neither superadditivity nor concavity encapsulate the meaning of the
second law in all situations.
\end{abstract}

Lower dimensional gravity continues to remain a subject of active research,
primarily due to the insights it might yield into the nature of quantum
gravity. In particular, although the Einstein tensor is trivially zero in
two spacetime dimensions, it has been shown that black holes do exist in
this setting \cite{MST,BHT}. The associated theory of gravitation from
which they arise is formed by setting the Ricci scalar equal to the trace
of the conserved stress energy tensor \cite{MFound}; despite its
simplicity, this theory also has a number of other striking classical and
semi-classical features, including a well-defined Newtonian limit
\cite{MFound}, a post-Newtonian expansion, FRW cosmologies, gravitational
collapse \cite{Arnold} and black hole radiation
\cite{TomRobb,2dsemi,Shardir}.  Indeed, in a certain sense the field
equations  are the $D\to 2$ limit of the $D$-dimensional Einstein equations
\cite{RSD2}. More recently, it has been shown that $(1+1)$-dimensional
black holes can also arise from string-motivated dilaton theories of
gravity in two dimensions \cite{MSW}.

Static black hole metrics in two spacetime dimensions are conveniently
exhibited in the form
\begin{equation}
ds^2 = - \alpha(x) dt^2 + \alpha^{-1}(x) dx^2   \label{01}
\end{equation}
where $\alpha(x)$ is determined from field equations which follow from
an action for two dimensional gravity whose general form we shall outline
below. The spacetime (\ref{01}) is that of a static $(1+1)$ dimensional
black hole if $\alpha(x)< 0$ for some spatial region $x\in (b,a)$.
Dimensional arguments \cite{TomRobb} indicate that the temperature
associated with such black holes is proportional to the ADM-mass M.
This quantity has units of inverse length, and so metrics of the form
(\ref{01}) depend upon $M$ as $\alpha = \alpha(Mx,Q_p/M^p)$, where $Q_p$
is a coupling constant that appears in the matter action of dimension
(length)$^{-p}$. Standard Wick-rotation arguments \cite{MST} yield
\be
T = \frac{M}{2\pi}\vert\frac{d\alpha(y,Q_p/M^p)}{dy}\vert_{y_H}
\label{01a}
\ee
where the horizon $x_H = y_H(M)/M$ is defined via $\alpha(y_H,Q_p/M^p) = 0$.
Provided that $\frac{d\alpha(y,Q_p/M^p)}{dy}$ is independent of $Q$
(a situation which holds in a large number of cases \cite{TomRobb,Rtdilat})
the entropy varies as $\ln(M/M_0)$, where $M_0 \neq 0$ is a constant of
integration which arises from integrating the thermodynamic relation
$dU = T dS$. As the black hole radiates away its mass its Compton
wavelength becomes comparable to its Schwarzchild radius, beyond which
the thermodynamic limit breaks down and a full quantum theory of gravity is
required. The zero of entropy may be chosen to be at this mass scale,
thereby defining the constant $M_0$.  We shall assume henceforth that this
situation holds.

The $(1+1)$-dimensional black hole gives rise to some interesting points
of principle in thermodynamics as will be shown in this paper. In order to
explain this we recall some relevant earlier results \cite{PF1,PF2,PTB}.
Let $X_a$,
$X_b$ be two values of a set of extensive variables referring to a specific
system like a gas, a solid, a black hole, {\it etc}. Then the entropy can
satisfy any of the following properties:
\begin{eqnarray}
\mbox{superadditivity} \quad\S :&\qquad \qquad \ \ \quad S(X_a+X_b) \ge&
S(X_a)+S(X_B) \nonumber \\
\mbox{homogeneity} \quad\H :&\qquad \qquad\qquad\ \quad S(\mu X_a) =& \mu
S(X_a) \label{1} \\
\mbox{concavity} \quad\C :&\qquad S(\lambda X_a+ (1-\lambda)X_b) \ge&
\lambda S(X_a)+ (1-\lambda) S(X_b) \nonumber
\end{eqnarray}
where $\mu$ and $\lambda$ ($0\le\lambda\le 1$) are arbitrary positive
constants.

$\S$ states that on withdrawing a partition so that two systems of a similar
type can merge, the entropy increases or remains unchanged. $\H$ asserts the
extensive nature of the entropy and $\C$ is its concavity.  Note that
\be
{\S} + {\H} \Rightarrow \C      \label{2}
\ee
which holds because $\S$ and $\H$ respectively imply $S(\lambda X
+ (1-\lambda) Y)\ge S(\lambda X)+S((1-\lambda) Y) = \lambda S(X)+ (1-\lambda)
S(Y)$, which is \C.
Similarly,
\be
{\C} + {\H} \Rightarrow {\S}   \label{3}
\ee
because $\C$ and $\H$ respectively imply $S(\lambda X  +(1-\lambda) Y)\ge
\lambda S(X)+ (1-\lambda) S(Y) = S(\lambda X)+S((1-\lambda) Y)$; setting
$\lambda X\equiv X_a$ and $(1-\lambda)Y\equiv X_b$ yields $\S$ .

As a consequence, of the eight conceivable types of thermodynamics
\begin{eqnarray}
{\S\H\C} & {\S\Hbar\C}\quad{\underline{\S\H\Cbar}} &
{\S\Hbar\Cbar} \nonumber \\
{\underline{\Sbar\H\C}} & {\Sbar\Hbar\C}\quad{\Sbar\H\Cbar} &
{\Sbar\Hbar\Cbar}  \label{4}
\end{eqnarray}
not all are logically admissible; we have underlined the inadmissible ones
in (\ref{4}) above. A third implication
\be
{\S} + {\C} \Rightarrow {\H}   \label{5}
\ee
may be proved subject to the assumption that $\S(0)=0$
\cite{PF3}. If (\ref{5}) holds, this would imply that  \be {\S\Hbar\C}
\label{6} \ee is not allowed.

Up to the present it was believed that $\S$ was a requirement, leaving only
the three types in the top line of (\ref{4}), and only two if (\ref{6}) is
excluded. Of these, $\S\H\C$ is indeed the normal type of thermodynamics,
while $\S\Hbar\Cbar$ applies to three-dimensional black holes of mass $M$
for which the entropy $S= \frac{4\pi G}{k} M^2$ ensures that $\H$ and $\C$
are violated. This then provided examples of apparently all available types
of thermodynamics.

For $(1+1)$-dimensional black holes, we will show that the
dimensionless entropy
\be
\sigma_0 \equiv \frac{2\pi}{\hbar k} S = \ln(\frac{M}{M_0}) \label{7}
\ee
satisfies $\S\Hbar\C$ for $M/M_0 < 4$ and $\Sbar\Hbar\C$ for $M/M_0 > 4$,
thus adding an example for each of two relatively unexplored types of
thermodynamics. It is the circumstance that $S(0)\neq 0$ that brings the
type (\ref{6}) into play.  The fact that $S(0)$ is negative is a key
characteristic of the MATHEMATICAL equation for $S(X)$. Note that therefore
its PHYSICAL applicability must cease just before the dashed portions of
the curves of fig. 1 are reached.

Consider a $(1+1)$-dimensional system consisting of a black hole of mass
$M$ which has spawned $n$ identical black holes of mass $m$. The
dimensionless entropy is
\be
\sigma_n = \ln\left[\frac{M-nm}{M_0}\left(\frac{m}{M_0}\right)^n\right]
\label{8}
\ee
provided the subsystems are sufficently separated
to be regarded as independent. Eq. (\ref{8}) becomes
(\ref{7}) for $n=0$. It is easily confirmed that for a given
$n$, $M$ and $M_0$, $\sigma_n$ has its largest value if $m=M/(n+1)$, whence
\be
\sigma^{\rm max}_n = (n+1)\ln\left[\frac{M}{(n+1)M_0}\right] \quad .
\label{9}
\ee
Fig. 1 shows a plot of $\sigma_1$ as a function of $m/M_0$ and also gives
an indication of $\sigma_0$.  The changes indicated by arrows in fig.1 must
occur because the second law states, for our purposes here and for many
other purposes as well, that the entropy increases,  or possibly stays
constant, in spontaneous processes. For $M/M_0=3$ the separated black holes
merge, and the system is of the type $\S\Hbar\C$; for $M/M_0=5$ the black
hole tends to split into equal parts and the system is of the type
$\Sbar\Hbar\C$. The concave nature of the logarithmic function ensures the
validity of $\C$ in both cases.

For $n$ large enough to permit differentiation of $\sigma^{\rm max}_n$
with respect
to $n$, one finds (given $M$ and $M_0$)
the maximal value of $\sigma^{\rm max}_n$ (say, $\hat{\sigma}_{n_0}$)
to be at $n=n_0$, where
\be
n_0 = \frac{M}{e M_0} -1 \quad\mbox{and}\quad
\hat{\sigma}_{n_0} = \frac{M}{e M_0} = n_0 + 1 \quad . \label{10}
\ee
Strictly speaking $n_0$ must be an integer and $\hat{\sigma}_{n_0}$ is a
rising staircase curve as a function of $n_0$.

To find the entropically  best transition for a black hole of mass M
one must ensure that $\sigma_n > \sigma_0$, or alternatively $M/M_0 >
(n+1)^{1+1/n}$. This yields the number of equal black holes into which the
original one will fragment.  Equating $\sigma_0$ and $\sigma_1$  yields a
curve of ``indifference'' when the black hole system can split up, or not,
without a change of entropy. This occurs for
\be
\frac{M}{M_0}=4 \qquad\mbox{where}\quad \sigma_0=\sigma_1 =2\ln(2)=1.386 \quad
{}.
\ee
This case is also illustrated in Fig. 1. For no fragmentation $M/M_0$ is less
than 4; so the fragmentation process always comes to a stop, as one
intuitively expects.

The entropy formula (\ref{7}) has been shown to follow from the static
black hole metric (\ref{01}) using both naive Wick rotation and more
rigorous quantum field theoretic arguments \cite{TomRobb}. Such metrics
arise from theories of $(1+1)$-dimensional gravity based on the action
\begin{equation}
S[g,\Psi]
= \int d^2x\sqrt{-g}\left( H(\Psi)g^{\mu\nu}\nabla_\mu\Psi
   \nabla_\nu\Psi +D(\Psi)R + V(\Psi;\Phi_M)\right) \label{1a}
\end{equation}
where $R$ is the Ricci scalar and $D$ and $H$ are arbitrary functions of
the scalar field $\Psi$ (referred to as the dilaton in string theory). The
potential $V$ is the matter Lagrangian, and depends on both $\Psi$ and the
matter fields $\Phi_M$; in general the associated conserved stress-energy
tensor $T_{\mu\nu}$ will depend upon both fields.
One need only require that static metrics which
are solutions to the field equations which follow from (\ref{1a})
have the asymptotic form
\be
\alpha(x) \rightarrow 2M |x| -1  \label{03}
\ee
for large $|x|$ since in this case the metric
(\ref{01}) will become asymptotically like a Rindler spacetime; a Rindler
transformation may then be applied locally to  rewrite the metric in
Minkowskian form. Collapse of $(1+1)$ dimensional dust \cite{Arnold}
has been shown to yield black hole metrics whose exterior metric matches
onto solutions with such asymptotic behaviour.

As an example of a specific black hole metric, if $D(\Psi)=\Psi$, $H=1/2$
and $V(\Phi)$ is the the stress-energy tensor of a Liouville field coupled
to gravity, then
\be
\alpha(x) = 1 - q e^{-2 M |x|}  \label{04}
\ee
where $\Lambda = q M^2$ is the coupling constant. The parameter $M$ is the
ADM-mass \cite{Rtdilat} and we have set $2\pi G =1$. Note that (\ref{04})
is also the black-hole metric of string theory \cite{MSW} if one
interchanges the parameters $M$ and $\sqrt{\Lambda}$; this yields radically
different thermodynamics from that which we are considering.  The metric
(\ref{04}) is asymptotically flat for large $|x|$, has an event horizon at
$|x|=\ln(q)/(2M)$, and a temperature $M/(2\pi)$. Our discussion here is
also pertinent to the case of $(2+1)$ dimensional black holes \cite{3dbh},
for which $T \sim \sqrt{M}$ and $S \sim \sqrt{M}$, again leading to a
breakdown of superadditivity.

We should explain that one really needs the {\it dynamics} of black holes
moving apart in order to explain how they can fragment. This is presumably
a hard GR problem. All one can be sure about is that the final state has
a higher entropy.

Can the essence of the second law of thermodynamics be conveyed by a
functional property of the entropy (in terms of key extensive variables
like energy, volume, mass etc.)? It has been maintained by some that it
resides in the superadditivity of the entropy and by others that it resides
in the concavity of the entropy.  We have shown here that neither position
can now be maintained universally. $\H$ and $\Hbar$, $\C$ and $\Cbar$ and
also $\S$ and $\Sbar$ can characterize the entropy. Our study of
$(1+1)$-dimensional black hole thermodynamics has doubled the number of
useful and relevant types of thermodynamics from two to four. This still
leaves one with a search for physically relevant examples for the types
$\Sbar\H\Cbar$ and $\Sbar\Hbar\Cbar$.

\section*{Acknowledgements}

In ref. \cite{PTB}, tables 5 and 6, we put $\Sbar$, $\Hbar$ $\Rightarrow$
$\C$ , when only $\Sbar$, $\Hbar$ should have been written, nor was the
assumption $S(0) = 0$ explicitly noted in \cite{PTB,PF3} for $\S + \C
\Rightarrow \H$. We are indebted to Dr. R. Woodard, University of Florida,
Gainesville  for enlightening discussions of the various types of
thermodynamics. This work was supported in part by the  Natural Sciences
and Engineering Research Council of Canada.

 \par\vfill\eject

\section*{Figure Captions}

\noindent
{\bf Fig. 1} The dimensionless entropy of a combination of two separated
$(1+1)$-dimensional black holes of masses $M-m$ and $m$ as a function of
$m/M_0$. For $M/M_0=3$ on has the $\S\Hbar\C$ type of thermodynamics and for
$M/M_0=5$  one has the $\Sbar\Hbar\C$ type. The curve $M/M_0=4$ represents
a limiting case.


\begin{thebibliography}{References}

\bibitem{MST}R.B. Mann, A. Shiekh, and L. Tarasov, Nucl. Phys. {\bf B341}
(1990) 134.
\bibitem{BHT}J.D. Brown, M. Henneaux and C. Teitelboim,
Phys. Rev. {\bf D33} (1986) 319; J.D. Brown, {\sl Lower
Dimensional Gravity}, (World Scientific, 1988).
\bibitem{MFound}R.B. Mann, Found. Phys. Lett. {\bf 4} (1991) 425.
\bibitem{Arnold}A.E. Sikkema and R.B. Mann, Class. Quantum
Grav. {\bf 8} (1991) 219.
\bibitem{TomRobb}R.B. Mann and T.G. Steele, Class. Quant. Grav. {\bf 9}
(1992) 475; R.B. Mann, L. Tarasov and A. Zelnikov, Class. Quant. Grav.
{\bf 9} (1992) 1487.
\bibitem{2dsemi}R.B. Mann, S.M. Morsink, A.E. Sikkema and T.G.
Steele, Phys. Rev. {\bf D43} (1991) 3948.
\bibitem{Shardir}S.M. Morsink and R.B. Mann, Class.
Quant. Grav. {\bf 8} (1991) 2257.
\bibitem{RSD2}R.B. Mann and S.F. Ross, Class. Quant. Grav. (to be published).
\bibitem{MSW}G. Mandal, A.M. Sengupta, and S.R. Wadia, Mod. Phys. Lett. {\bf
6} (1991) 1685.
\bibitem{Rtdilat}R.B. Mann, Phys. Rev. {\bf D47} (1993) 4438.
\bibitem{PF1}P.T. Landsberg and D. Tranah, Collective Phenomena {\bf 3}
(1980) 73.
\bibitem{PF2}D. Tranah and P.T. Landsberg, Collective Phenomena {\bf 3}
(1980) 81.
\bibitem{PTB}P.T. Landsberg in {\sl Black Hole Physics}, eds. V. de Sabbata
and Z. Zhang, (Dordrecht, Kluver 1992).
\bibitem{PF3}P.T. Landsberg and D. Tranah, Phys. Lett. {\bf A78} (1980)
219.
\bibitem{3dbh}M.~Ba\~nados, C.~Teitelboim and J.~Zanelli, Phys.
Rev. Lett. {\bf 69}, 1849 (1992); R.B. Mann and S.F. Ross,
Phys. Rev. {\bf D47} (1993) 3319.

\end{thebibliography}
\end{document}